# Non-resonant microwave absorption studies on MgB$_2$ thin film

S.P.Chockalingam, S.Sarangi, Ajay Sharma and S.V.Bhat*

Department of Physics, Indian Institute of Science, Bangalore-560012, India

[*]Corresponding author:

   Prof. S.V.Bhat

   Department of Physics

   Indian Institute of Science

   Bangalore –560012, India

   e-mail: svbhat@physics.iisc.ernet.in

   Tel.: +91-80-22932727. Fax: +91-80-23602602

**Abstract:**

We report on the presence of weak links in the intermetallic superconductor $MgB_2$ from the non-resonant microwave absorption (NRMA) studies on the $MgB_2$ thin film. The strong low field signal, which is the characteristics of weak links and the presence of hysteresis, which arises due to flux trapping, are observed. The studies are carried out as a function of temperature, microwave power, magnitude of modulation field and the range of field scan. Presence of weak links is discussed in terms of increase in the oxygen content of grain boundaries in the thin film and because of disorder effects arises due to the aging of the sample. The pinning of flux lines with the changes in sweeping direction leads to hysteresis in the NRMA signal and shows the contribution of flux line motion in the power absorption.

# Introduction:

The role of weak links in superconductors has been studied for a long time. Understanding the behavior of weak links has great importance for the implications on the application potential of superconductors. Presence of weak links in high $T_c$ materials due to its short coherence length and insulating grain boundaries limits the application potential of those materials. These weak links leads to lower critical current density and lower critical field of superconductors and leads to losses[1,2,3].

The discovery of superconductivity in the simple intermetallic compound $MgB_2$ has created a lot of interest from both application aspect and of fundamental science. Apart from its low cost and simple structure the technological importance of the sample arises due to its high critical current densities of the order of $10^6$ A/cm$^2$ and high upper critical magnetic field, which are favorable for large-scale applications and electronic devices. These properties which make $MgB_2$ differ from high $T_c$ materials as a potential candidate for applications and it has been suggested they arises due to the absence of weak links in $MgB_2$. The absence of weak links is explained in terms of its large coherence length, which would exceed narrow boundary widths, so that grain boundaries are highly transparent to supercurrent. Later it has been observed that in many cases grain boundary exceeds the coherence length, but transparency of supercurrent occurs due to the metallic behavior of grain boundaries, which act as proximity coupled SNS Josephson junction of high conductivity not like SIS weak link in high $T_c$ material[4,5]. Absence of weak links is reported in most of the $MgB_2$ literatures [6-11]only in very few studies presence of weak links is reported. These studies revealed the presence of weak links through the observation of presence of even harmonics generation response when it is subjected to ac and dc magnetic field [12] voltage flux oscillations in rf squid effects [13] vertical step in resonance curves from coplanar microwave resonator technique [14] second harmonic frequency emission at microwave frequency [15] and staircase response in I-V characteristics [16]. In this paper we reported the results of non-resonant microwave absorption (NRMA) studies on the $MgB_2$ thin film for the first time. The studies on $MgB_2$ thin film is of great importance from the application perspective since the thin films play a vital role in device making.

NRMA is a highly sensitive, non-invasive technique, which has proven to be a valuable tool for detecting weak links in superconductors and its characterization[17,18,19]. In this technique the sample is studied using a continuous wave electron paramagnetic resonance (EPR) spectrometer, by recording the magnetic field dependence of the power absorption. The magnetic field dependent signal from the sample is then compared with the modulating ac field so as to detect only that component of the signal, which appears with the same frequency, and phase as the modulating field. Because of this frequency as well as phase sensitive detection, a large signal to noise ratio is obtained and make this technique as an extremely sensitive one. However the recorded signal is the derivative of the actual power absorbed.

The NRMA studies on the $MgB_2$ thin film shows the presence of weak links and hysteresis in the signal. The origin of weak links is discussed due to increase in the oxygen content of the some grain boundaries due to oxidation and the effects of disorder arises due to the aging of the sample. The hysteresis appears because of pinning of flux lines when there is the change in the sweeping field direction. The NRMA studies are carried out as a function of temperature, modulation field, microwave power and the scan range and the results are reported.

## Experimental:

MgB$_2$ thin films were deposited on the sapphire substrate by PLD method as discussed in earlier reports [20]. A rectangular film of size 4 mm x 6 mm of typically 1000 nm thick was chosen for the NRMA studies. The XRD shows that the film is highly oriented along MgB$_2$ (002) as shown in the fig. 1(a). The film is highly homogenous with the grain sizes around 200 nm as seen in the SEM image of the film as in fig. 1(b) The film showed a zero resistance transition temperature of 39K.

The NRMA signals of the thin film was recorded using Bruker XPS X band (microwave frequency: 9.43 GHz) EPR spectrometer equipped with oxford Instruments ESR 900 continuous flow cryostat, in the temperature range of 4.2 K to room temperature. The NRMA measurements are done by the so-called homodyne lock in detection or phase sensitive detection. The measurements are done with the dc field applied parallel to the c-axis of the thin film. Unless otherwise specified, for the NRMA measurements following parameters are used. The modulation field of 4 Gauss peak to peak at 100 kHz, which is high enough so as to minimize the low modulation field induced hysteresis effects [21]. The signals are recorded between –50 to 150 gauss for both forward and reverse scans of the field with the microwave power of 21.01 mW. All the measurements are carried out with external dc applied parallel to the c-axis of the film and signals are recorded during both cooling and warming of the sample, which are essentially similar, and in this report the results during warming of the sample is presented.

## Results:

Fig.2 (a) shows the NRMA spectrum of $MgB_2$ film at different temperatures for both forward and reverse scans. In the normal state no signal is observed and a giant zero field minimum signal is observed in the superconducting state. At 38 K the signal starts appearing and its intensity increases with decreasing temperatures and at low temperatures the signal diminishes as shown in the fig.2 (a). The intensity of the forward and reverse scan signals are same but they are not centered around zero field. Another important feature observed in the measurements is the presence of hysteresis between the forward and reverse scans of the magnetic field, which decreases with increasing temperature as shown in the fig.2 (b).

The results of NRMA studies carried out with different microwave powers at 27K are shown in the fig.3 (a). With the increasing power the hysteresis between the forward and reverse scans decreases linearly as in the fig.3 (b)

Fig.4 (a) shows the NRMA signals obtained at different modulation fields applied at 100 kHz at 27 K. The results show the increase of signal intensity with the increasing modulation field and the decrease in hysteresis, fig.4 (b) shows its dependence.

To find out the effect of scanning field on the NMRA signals we carried out measurements with symmetric scanning at different modulation fields and also scanning at different field ranges. In the symmetric scanning both forward and reverse scan signals are identical and centered around zero field but in the asymmetric scanning signals are not centered around zero field and also reverse scan signal has more width than the forward one. With increasing modulation field both the forward and reverse scan signals position shifted linearly towards zero field but the shift in the reverse scan signal is more in the asymmetric scanning as shown in the fig. 5(a). In the symmetric scanning both forward and reverse scan signals shifted in the identical manner as in 5(b). Fig.6 (a) shows the NRMA signals observed when the scanning field ranges is varied. The hysteresis between the forward and reverse scan signals increases with the increasing field range as shown in the fig.5 (b). There are few differences between the symmetrical and asymmetrical scanning signals in the hysteresis, signal width and the signal position in the scale of field.

Understanding the above NRMA results on the $MgB_2$ film is important from both application and scientific views. The results are explained in terms of presence of weak links and the trapping of flux lines.

## Discussion:

In superconductors there are two dominant processes, which contribute to NRMA at low fields. One the decoupling of Josephson junctions due to magnetic fields which results in giant sharp width signals without hysteresis and the other the Lorentz force driven motion of quantized flux lines which shows a large hysteresis [9,10]. In our NRMA studies on $MgB_2$ thin film we observed both giant signal and hysteresis, which shows the presence of both weak links and flux trapping in the $MgB_2$ film that we studied.

Though there are many papers reported about the absence of weak links in $MgB_2$, few recent reports shows the presence of weak links in $MgB_2$ under certain conditions. Earlier NRMA studies on the bulk $MgB_2$ show the absence of weak links[8,9,10]. For the first time we are reporting the NRMA studies on the $MgB_2$ thin film, which confirmed the presence of weak links. The giant NRMA signal observed in the superconducting state is the characteristics of the weak links, which was earlier observed in high $T_c$ superconductors [17-19]. In the absence of weak links when the microwave power is applied to the sample there will be only little absorption due to flux motion and the absence of strong signals. But when the weak links are present the applied microwave power decoupled the weak links which act as josephson junctions in to the normal state and the absorption takes place which leads to a giant signal. In our studies on the thin film we observed this signal just like in high $T_c$, which is surprising, and confirm the presence of weak links in $MgB_2$ thin film.

The possibility of presence of weak links in $MgB_2$ arises due to various factors like the processing conditions, irradiation effects, sample morphology and purity and the disorder due to aging effects as discussed in the recent reports [12-16]. Our observation of weak links in $MgB_2$ thin film reveals that oxygen content at the grain boundaries plays a major role as discussed in the earlier report[22]. The SEM image of the thin film of dimension 4 mm x 6 mm which are 6 months old in which we did the NRMA studies shows the homogenous deposition of grains whose size is distributed in the range of 200-300nm. Due to the smaller grain sizes large no. of grain boundaries are present and they dominate in the thin film characteristics. These aged films when they are exposed to the atmosphere the oxygen content at the boundaries increased and some of them act as weak links in the intermetallic $MgB_2$ [22]. Meanwhile disorder effects arise due to the aging of the sample could not be ruled out. The NRMA technique is highly accessible to the surface of the sample, when this measurements is carried out in the thin film where surface properties dominates and exposed more grain boundaries as in the present sample, then the entire measurements turns out to extremely sensitive and the presence of weak links is observed.

The NRMA studies on $MgB_2$ film at different temperature are shown in the fig 2(a). At higher temperatures just below $T_c$ the no. of weak links is less so that the signal intensity.

With the decreasing temperature the weak links that act as Josephson junctions increases and they decoupled and leads to a giant signal. At lower temperatures the strength of the JJ increases and the weak links turns in to strong links and the signal intensity decreases. The hysteresis between the forward and reverse scans arises because of flux trapping [23,24], due to the changes in the sweeping field direction. The magnetic pressure gradient, which drives the fluxons, is removed when the direction of the field changes and fluxons get pinned till the reverse pressure gradient appears due to the field increased on the opposite side [25]. This pinning of fluxons is the cause for the presence of hysteresis between the forward and reverse scan. The increase in the temperature reduces the amount of flux trapped and hysteresis. The signals are not centered around zero field because of asymmetric field scanning ranges (-50 g to 150g) from the zero field.

The role of microwave power on the NRNA spectrum of $MgB_2$ film is studied by varying the applied power as show in the fig. 3(a). Since the measurements are carried out with the dc field applied parallel to the c-axis of the film, the microwave-induced currents on the sample surface are perpendicular to the dc field and the Lorentz force acting on the flux lines is maximum. Hysteresis decreases linearly with the increasing power as in the fig.3 (b). When the microwave power is increased the flux line motion also increases due to the increase in Lorentz force, which leads to more depinning and increased flux motion, which leads to more absorption. Also the increase in the microwave power decouples even stronger links if any with more coupling strength and leads to more loss.

Fig.4 (a) shows the effect of modulation field on the NRMA signals of the film at 27K. The presence of modulation field sweeps fluxons in and out of the Josephson junctions that increase the absorption with the increasing mod.field [26]. Hysteresis, which appears due to pinning of fluxons during the change in sweeping field direction, decreases with increasing modulation field, which provides pressure gradient and depinned the fluxons as shown in 3(b).

In asymmetric scanning signals are not centered around zero field, because flux trapped in the two directions are different due to the difference in the amplitude of the sweeping field range. But in symmetric scanning signals are centered around zero field because of equal amount of flux trapped in the two directions. The shift of the signals towards zero field with increasing modulation fields as in fig. 5(a) and 5(b) for symmetric and asymmetric scanning indicates the decrease in hysteresis as discussed above. Fig. 6 (a) shows the NRMA signals obtained for different sweeping fields, with the increasing field the reverse scan signals shifted away from the zero field. Since more fluxons get trapped at higher magnetic fields [27], the hysteresis increases linearly with sweeping field as in fig.6 (b). The signal width indicates the rate of power absorption with the field, more flux trapped then the signal width increases, if there is no flux trapping then the narrow signals are observed.

**Summary**:

For the first time we reported non-resonant microwave absorption (NRMA) studies on the MgB$_2$ thin film. The studies shows the presence of weak links, which act as Josephson junctions and dissipates the power when decoupled, another mechanism which contribute to the NRMA signal is motion of fluxons that also observed. The NRMA studies at different temperatures, modulation fields, microwave powers and sweeping fields are discussed

**Acknowledgements**, The authors thanks Prof.W.K.Kwok, Materials Science Division, Argonne National Laboratory, Argonne, Illinois and Prof N.Chandrasekhar, IMRE, Singapore for the MgB2 thin films, SVB thanks the University Grants Commission, India for funding this work.

.

# Figure Captions:

Figure 1(a): The XRD pattern of the MgB2 thin film, which shows that the film is highly oriented along (002).

Figure 1(b): The SEM image of the MgB2 thin film, which shows that the film is homogenously deposited with grain sizes distributed between 200- 300 nm.

Figure 2(a): NRMA signals obtained at different temperatures. To be noted are the presence of zero field minimum giant signals and the hysteresis between the forward and reverse scans. The signal starts appearing at temperatures closure to $T_c$ and then increases with decreasing temperature and again diminishes at lower temperatures.

Figure 2(b): Hysteresis between forward and reverse scans, as a function of temperature, To be noted is the decrease of hysteresis with increasing temperature, at higher temperature it decreases linearly.

Figure 3(a): NRMA signals resulted at different applied microwave power. To be noted are the increase in the intensity of the signal and the decrease in the hysteresis with the increasing power.

Figure 3(b): Hysteresis between forward and reverse scans, as a function of applied microwave power, To be noted is the linear decrease of hysteresis with increasing power.

Figure 4(a): NRMA signals obtained at different modulation field. To be noted are the increase in the signal intensity and the decrease in the hysteresis with the increasing modulation field applied at 100 KHz.

Figure 4(b): Hysteresis between forward and reverse scans as a function modulation field; To be noted is the linear decrease of hysteresis with increasing modulation field.

Figure 5(a): Shift in the NRMA signal measured from zero field as a function of different modulation field for asymmetric sweeping field. To be noted is the linear dependence of signal shift with modulation field.

Figure 5(b): Shift in the NRMA signal measured from zero field as a function of different modulation field for symmetric sweeping field. To be noted is the linear dependence of signal shift with modulation field.

Figure 6(a): NRMA signals obtained for different sweeping field ranges. To be noted are increase in the hysteresis, the increase in the width of the signal and its shift towards higher fields with the increasing sweeping field range.

Figure 6(b): Hysteresis between forward and reverse scans, as a function of sweeping field range, To be noted is the linear increase of hysteresis with increasing sweeping field amplitude.

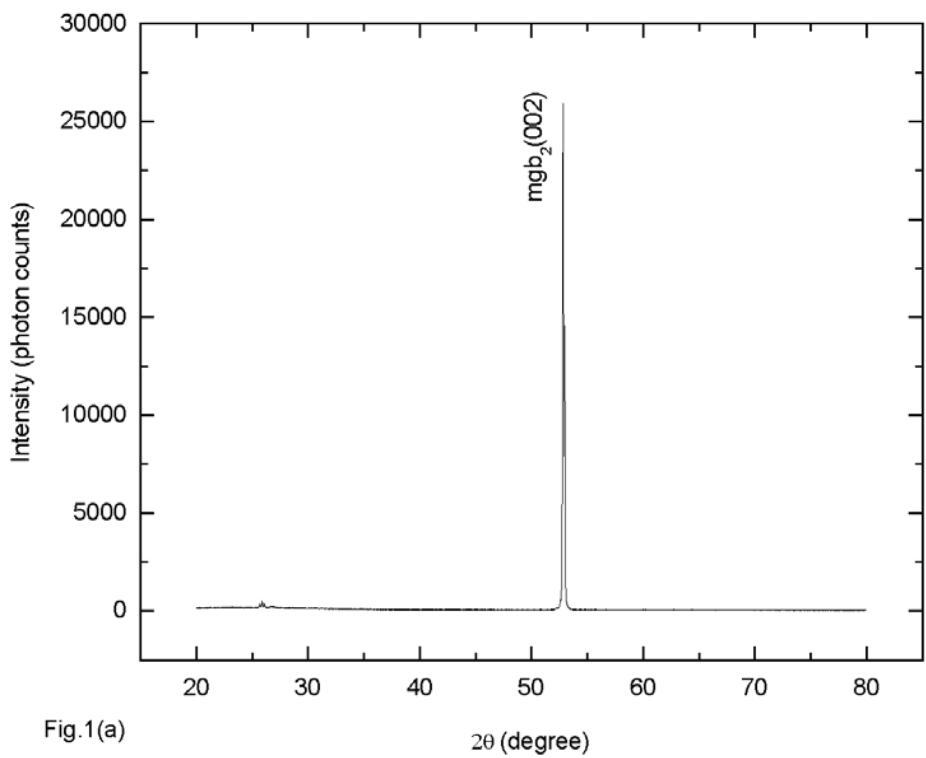

Fig.1(a)    2θ (degree)

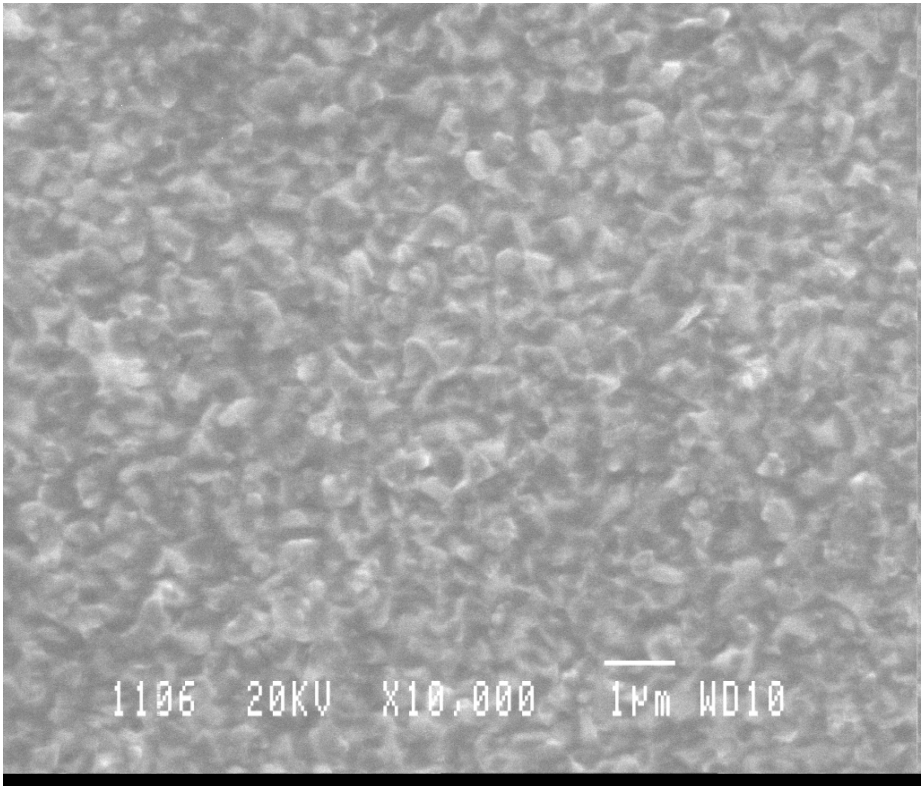

Fig. 1(b)

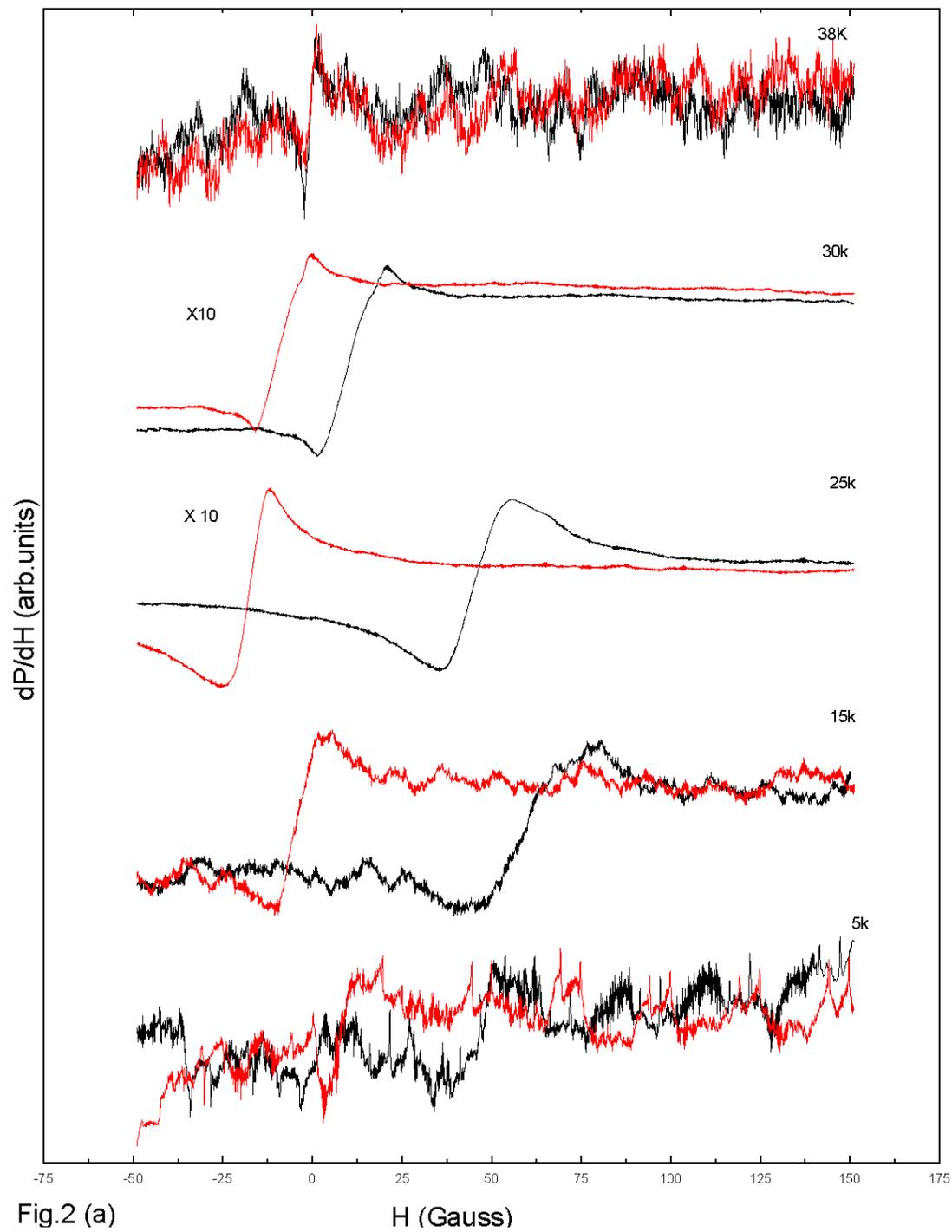

Fig.2 (a)

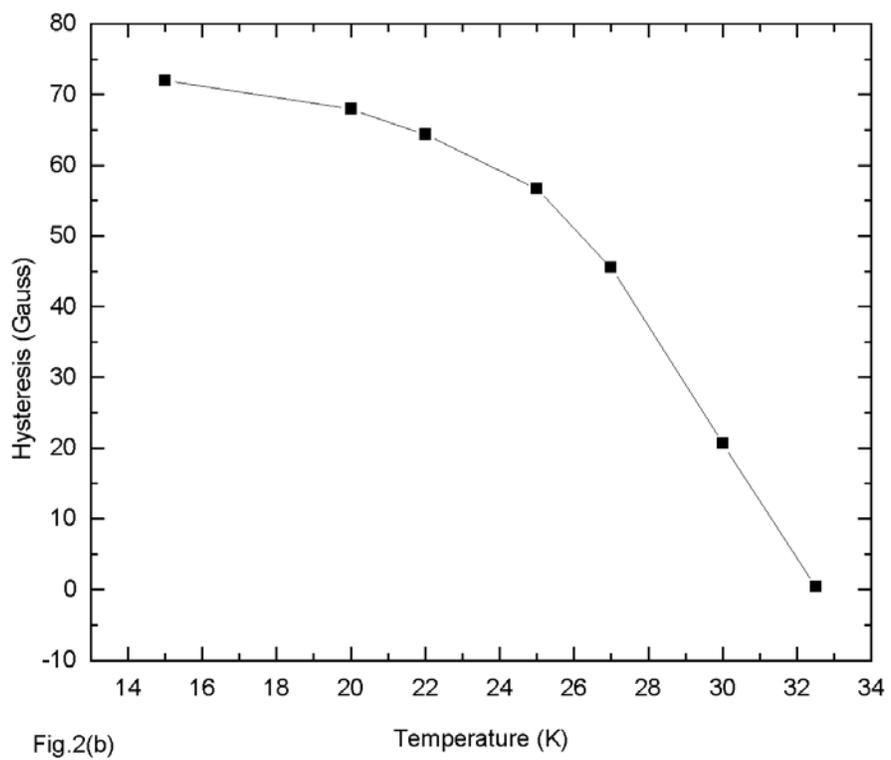
Fig.2(b)

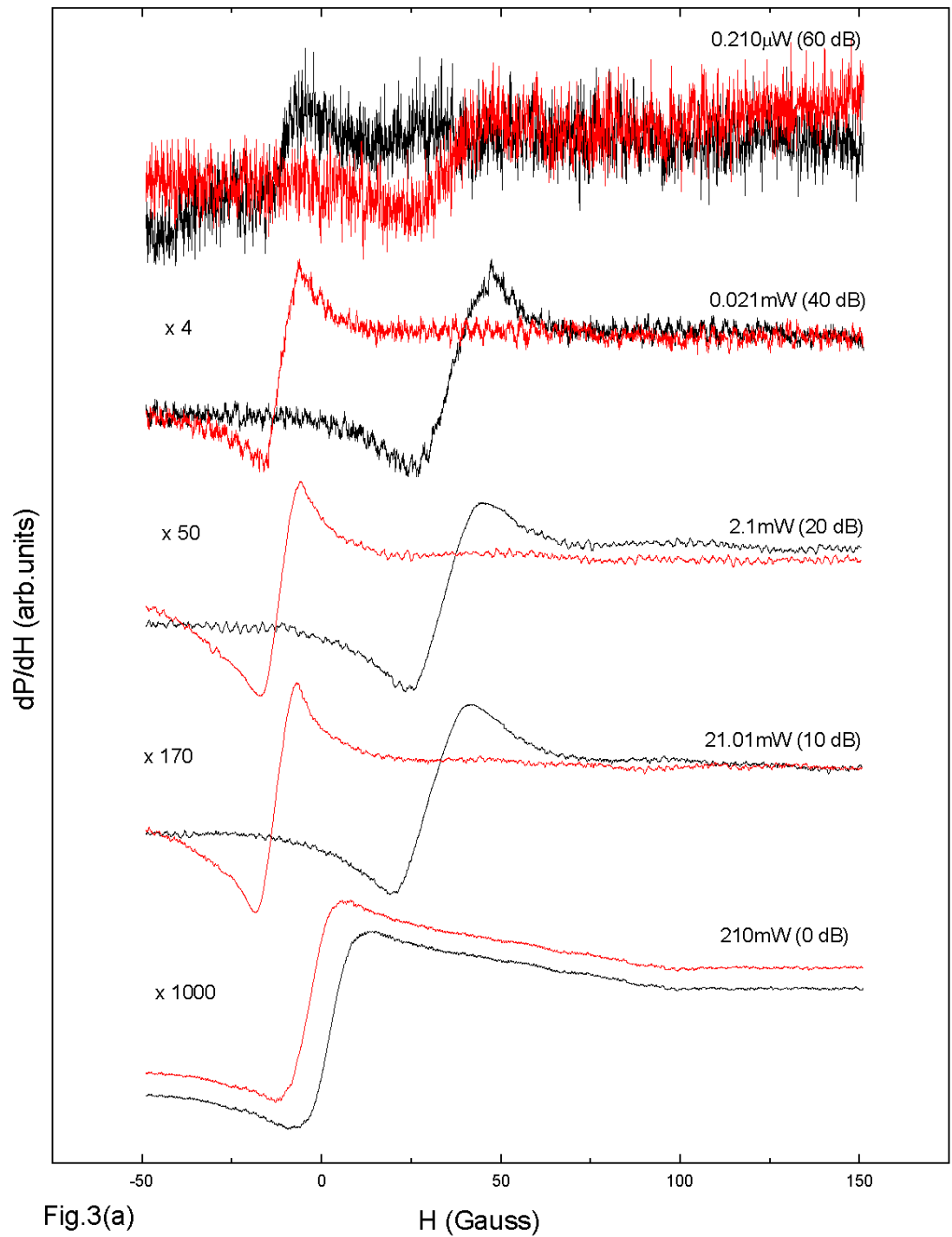

Fig.3(a)

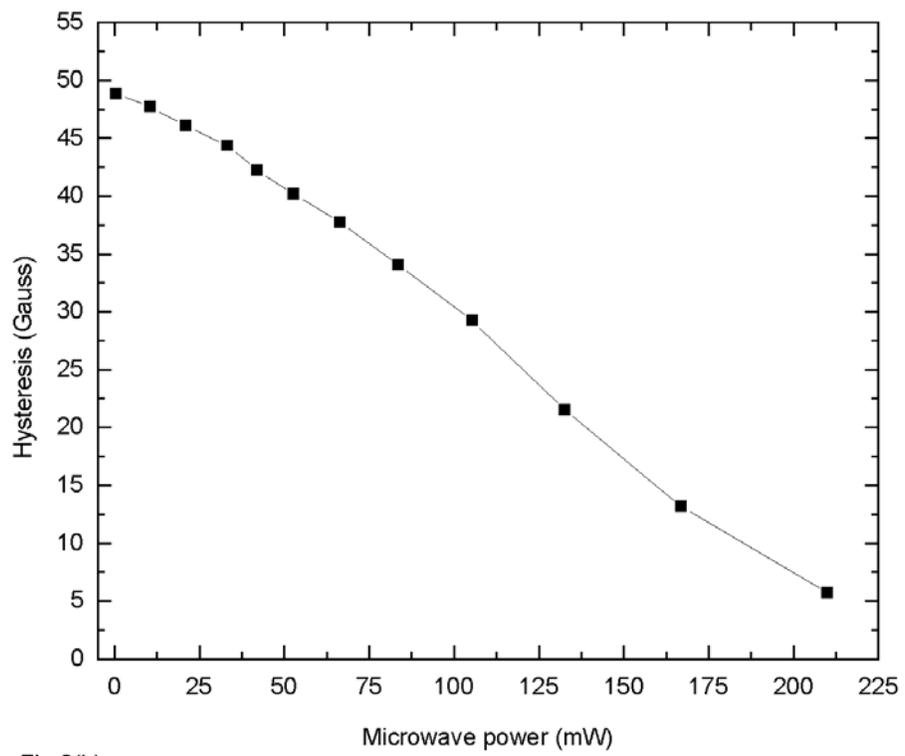

Fig.3(b)

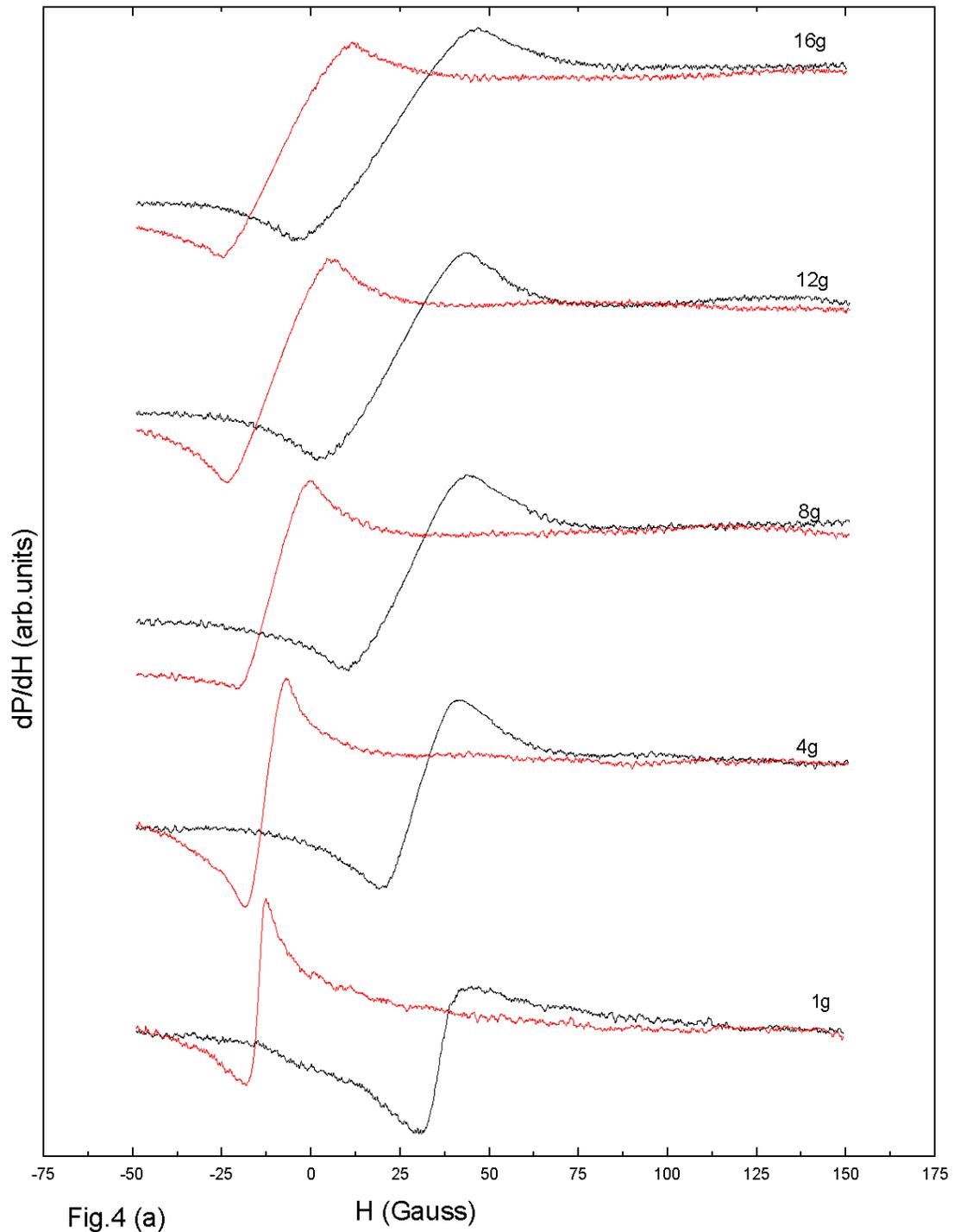

Fig.4 (a)

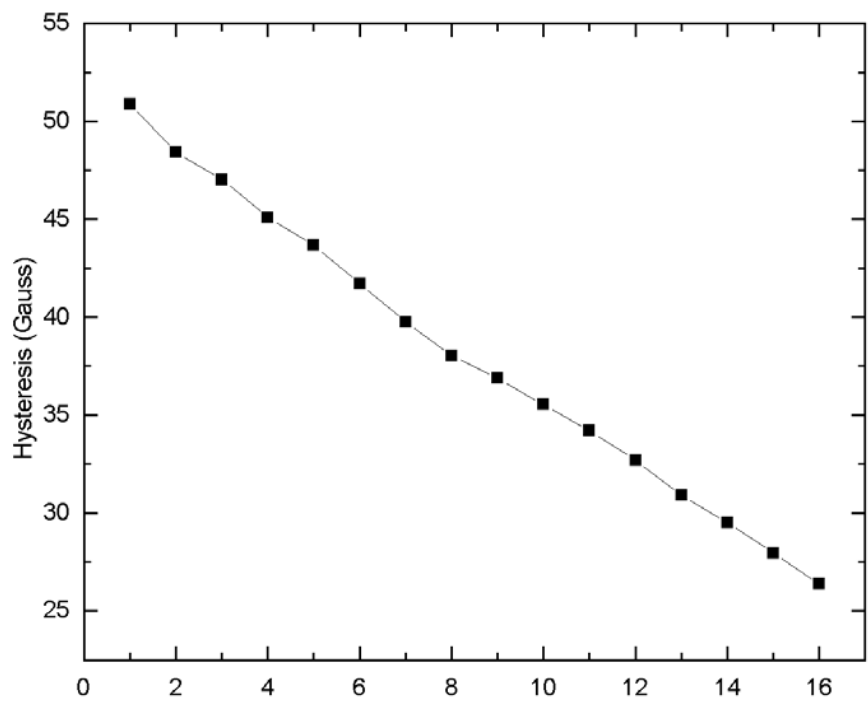

Fig.4 (b)

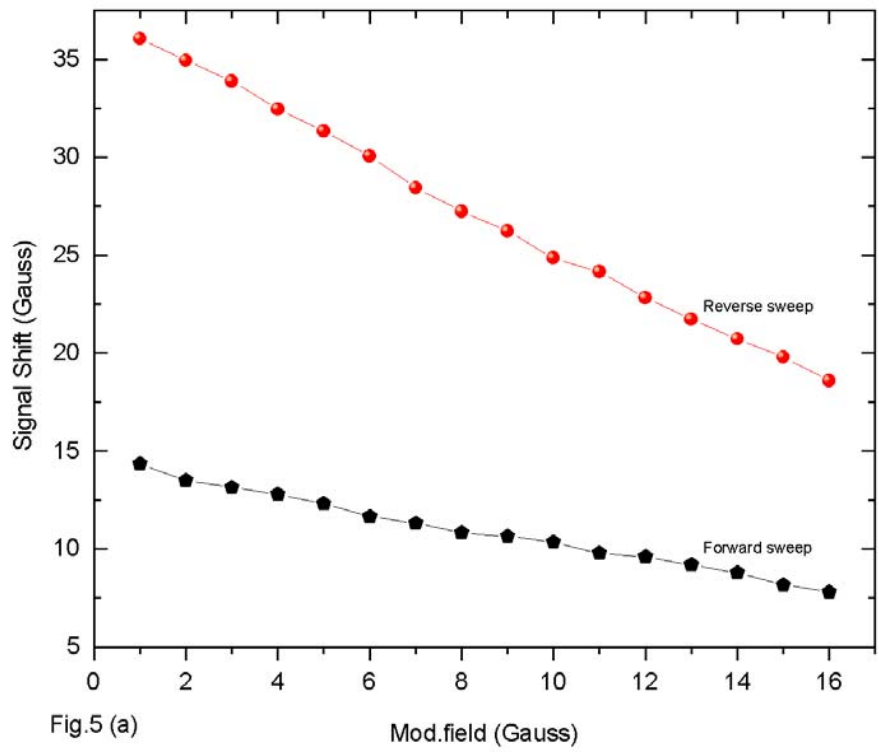

Fig.5 (a)

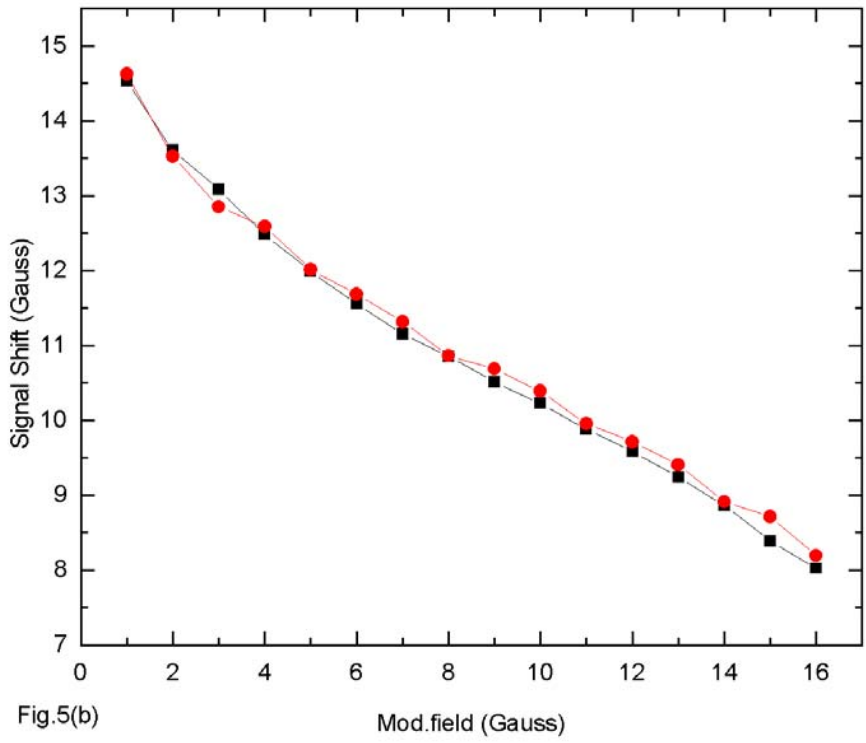
Fig.5(b)
24

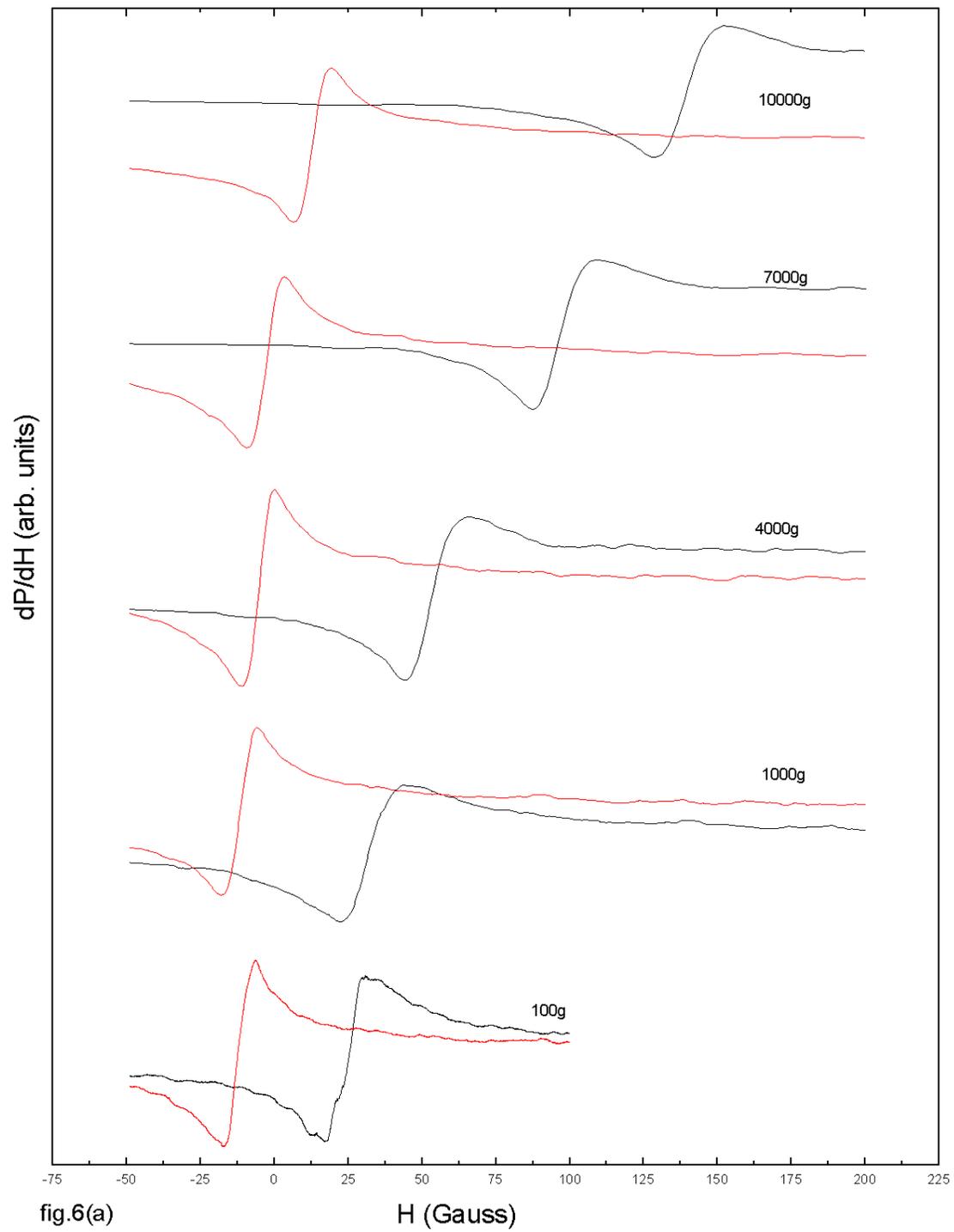

fig.6(a)

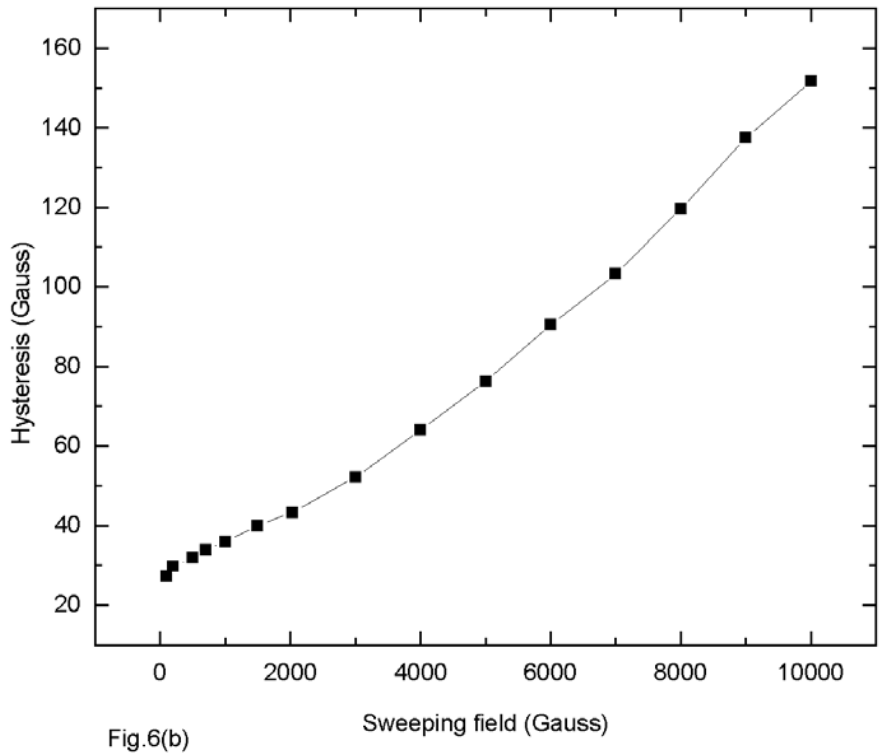

Fig.6(b)